\documentclass[11pt, twoside]{article}
\usepackage{fullpage}
\usepackage{amssymb,amsmath,amsthm}
\usepackage[dvips]{graphics}
\usepackage{epsfig}
\usepackage{color}
\usepackage[colorlinks]{hyperref}            
\usepackage{psfrag}
\usepackage[square,authoryear]{natbib}
\usepackage{floatflt}

\begin{document}
\newcommand{\todo}[1]{\vspace{5 mm}\par \noindent
\framebox{\begin{minipage}[c]{0.9 \textwidth}
\tt #1 \end{minipage}}\vspace{5 mm}\par}

\newcommand{\bfi}{\bfseries\itshape}
\renewcommand{\proofname}{\bf Proof}

\newcommand\nudgeback{\kern-4pt}
\gdef\MissingItem#1{\textcolor{blue}{#1}}  

\definecolor{dullmagenta}{RGB}{102,0,102}   
\definecolor{darkblue}{RGB}{0,0,153}   

\hypersetup{urlcolor=dullmagenta
           ,citecolor=dullmagenta
           ,linkcolor=dullmagenta
           ,pagecolor=white
           }

\noindent

\newcommand{\AIAAJ}{AIAA J.}
\newcommand{\AIAAP}{AIAA Paper}
\newcommand{\AM}{Acta Math.}
\newcommand{\ARMA}{Archive for Rational Mechanics and Analysis}
\newcommand{\ASMEJFE}{J. Fluids Eng., Trans. ASME}
\newcommand{\ASR}{Applied Scientific Research}
\newcommand{\CF}{Computers Fluids}
\newcommand{\ETFS}{Experimental Thermal and Fluid Science}
\newcommand{\EF}{Experiments in Fluids}
\newcommand{\FDR}{Fluid Dynamics Research}
\newcommand{\IJHMT}{Int. J. Heat Mass Transfer}
\newcommand{\JASA}{J. Acoust. Soc. Am.}
\newcommand{\JCP}{J. Comp. Physics}
\newcommand{\JFM}{J. Fluid Mech}
\newcommand{\JMP}{J. Math. Phys.}
\newcommand{\JSC}{J. Scientific Computing}
\newcommand{\JSP}{J. Stat. Phys.}
\newcommand{\JSV}{J. of Sound and Vibration}
\newcommand{\MC}{Mathematics of Computation}
\newcommand{\MWR}{Monthly Weather Review}
\newcommand{\PAS}{Prog. in Aerospace. Sci.}
\newcommand{\PCPS}{Proc. Camb. Phil. Soc.}
\newcommand{\PD}{Physica D}
\newcommand{\PRA}{Physical Rev. A}
\newcommand{\PRE}{Physical Rev. E}
\newcommand{\PRL}{Phys. Rev. Lett.}
\newcommand{\PF}{Phys. Fluids}
\newcommand{\PFA}{Phys. Fluids A.}
\newcommand{\PL}{Phys. Lett.}
\newcommand{\PRSLA}{Proc. R. Soc. Lond. A}
\newcommand{\SIAMJMA}{SIAM J. Math. Anal.}
\newcommand{\SIAMJNA}{SIAM J. Numer. Anal.}
\newcommand{\SIAMJSC}{SIAM J. Sci. Comput.}
\newcommand{\SIAMJSSC}{SIAM J. Sci. Stat. Comput.}
\newcommand{\TCFD}{Theoret. Comput. Fluid Dynamics}
\newcommand{\ZAMM}{ZAMM}
\newcommand{\ZAMP}{ZAMP}

\newcommand{\ICASER}{ICASE Rep. No.}
\newcommand{\NASACR}{NASA CR}
\newcommand{\NASATM}{NASA TM}
\newcommand{\NASATP}{NASA TP}
\newcommand{\ARFM}{Ann. Rev. Fluid Mech.}
\newcommand{\WWW}{from {\tt www}.}
\newcommand{\CTR}{Center for Turbulence Research, Annual Research Briefs}
\newcommand{\vonKarman}{von Karman Institute for Fluid Dynamics Lecture Series}

\def\p{\partial}
\def\({\left(}
\def\){\right)}
\def\etal{{\it et al.\ }}
\def\ie{{\it i.e.\ }}
\def\eg{{\it e.g.\ }}

\def\Exponent{{-\beta\left[ 2\pi\sigma \( \overline{\psi}+y U \) + 
\alpha(\sigma) \right]}}

\def\tu{{\tilde u}}
\def\ta{{\tilde \alpha}}


\newcommand{\alphahb}{\widehat{\overline{\alpha}}}
\newcommand{\alphab}{\overline{\alpha}}
\newcommand{\alphah}{\widehat{\alpha}}
\newcommand{\uhb}{\widehat{\overline{u}}}
\newcommand{\ub}{\overline{u}}
\newcommand{\uh}{\widehat{u}}

\def \bu {\bar{u}}
\def \hu {\hat{\bar{u}}}


\title{ A Dynamic model for the \\
 Lagrangian Averaged Navier-Stokes-$\alpha$ Equations \vspace{0mm}} 
\author{
%
  \begin{minipage}{0.6\linewidth}
  \begin{center}
  Hongwu Zhao and Kamran Mohseni\\[1.0ex]
  Aerospace Engineering Sciences \\
  University of Colorado, 107-81 \\
  Boulder, CO 80309-0429 \\
  Tel: (303) 492 0286 (Mohseni)\\
  Fax: (303) 492 7881\\
  Email: {\tt mohseni@colorado.edu}
  \end{center}
  \end{minipage}
}
\maketitle

\section*{Abstract}

A {\it dynamic} procedure for the Lagrangian Averaged
Navier-Stokes-$\alpha$ (LANS-$\alpha$) equations is developed where
the variation in the parameter $\alpha$ in the direction of anisotropy
is determined in a self-consistent way from data contained in the
simulation itself.  In order to derive this model, the incompressible
Navier-Stokes equations are Helmholtz-filtered at the grid and a test
filter levels. A Germano type identity is derived by comparing the
filtered subgrid scale stress terms with those given in the
LANS-$\alpha$ equations. Assuming constant $\alpha$ in homogenous
directions of the flow and averaging in these directions, results in a
nonlinear equation for the parameter $\alpha$, which determines the
variation of $\alpha$ in the non-homogeneous directions or in
time. Consequently, the parameter $\alpha$ is calculated during the
simulation instead of a pre-defined value.  The dynamic model is
initially tested in forced and decaying isotropic turbulent flows
where $\alpha$ is constant in space but it is allowed to vary in time.
It is observed that by using the dynamic LANS-$\alpha$ procedure a
more accurate simulation of the isotropic homogeneous turbulence is
achieved. The energy spectra and the total kinetic energy decay are
captured more accurately as compared with the LANS-$\alpha$
simulations using a fixed $\alpha$. In order to evaluate the
applicability of the dynamic LANS-$\alpha$ model in anisotropic
turbulence, {\it a priori} test of a turbulent channel flow is
performed. It is found that the parameter $\alpha$ changes in the wall
normal direction. Near a solid wall, the length scale $\alpha$ is seen
to depend on the distance from the wall with a vanishing value at the
wall. On the other hand, away from the wall, where the turbulence is
more isotropic, $\alpha$ approaches an almost constant value.
Furthermore, the behavior of the subgrid scale stresses in the near
wall region is captured accurately by the dynamic LANS-$\alpha$
model. The dynamic LANS-$\alpha$ model has the potential to extend the
applicability of the LANS-$\alpha$ equations to more complicated
anisotropic flows.

\newpage
\section{Introduction}

Turbulent flows play a an important role in many areas of engineering
fluid mechanics as well as atmospheric and oceanic flows.  Accurate
simulation of a turbulent flow requires that the energetics of the
large scale energy containing eddies, dissipative small scales, and
inter-scale interactions to be accounted for. In direct numerical
simulations (DNS) all the involved scales are directly calculated. DNS
is believed to provide the most comprehensive representation of the
governing equations of fluid flows; the so-called Navier-Stokes (NS)
equations. Owing to the very high Reynolds numbers encountered in most
problems of interest, the disparity between the large scales and small
scales, which represents the computational size of the problem,
rapidly grows with the Reynolds number. Consequently, DNS can resolve
only a small fraction of the turbulent activity for high Reynolds
number flows.

While the direct numerical simulation of most engineering flows seems
unlikely in near future, turbulence modeling could provide qualitative
and in some cases quantitative measures for many applications. Large
Eddy Simulations (LES) and the Reynolds Averaged Navier-Stokes
Equations (RANS) are among the numerical techniques to reduce the
computational intensity of turbulent calculations. In LES, the
dynamics of the large turbulence length scales are simulated
accurately and the small scales are modeled. The vast majority of
contemporary LES make use of eddy-viscosity based Subgrid-Scale (SGS)
models in conjunction with the spatially-averaged (filtered)
Navier-Stokes Equations.  In this approach, the effect of the
unresolved turbulence is modeled as an effective increase in the
molecular viscosity.  
On the other hand, RANS models are obtained by time averaging the
Navier-Stokes equations. In this case most of the unsteadiness is
averaged out. Consequently, the time mean quantities are calculated
while the faster scale dynamics are modeled. RANS simulations are
often more affordable than LES, however, their accuracy is somewhat
limited in many applications \cite{Wilcox:93a}.
 
\begin{figure}
\begin{center}
\includegraphics[width=0.7\linewidth]{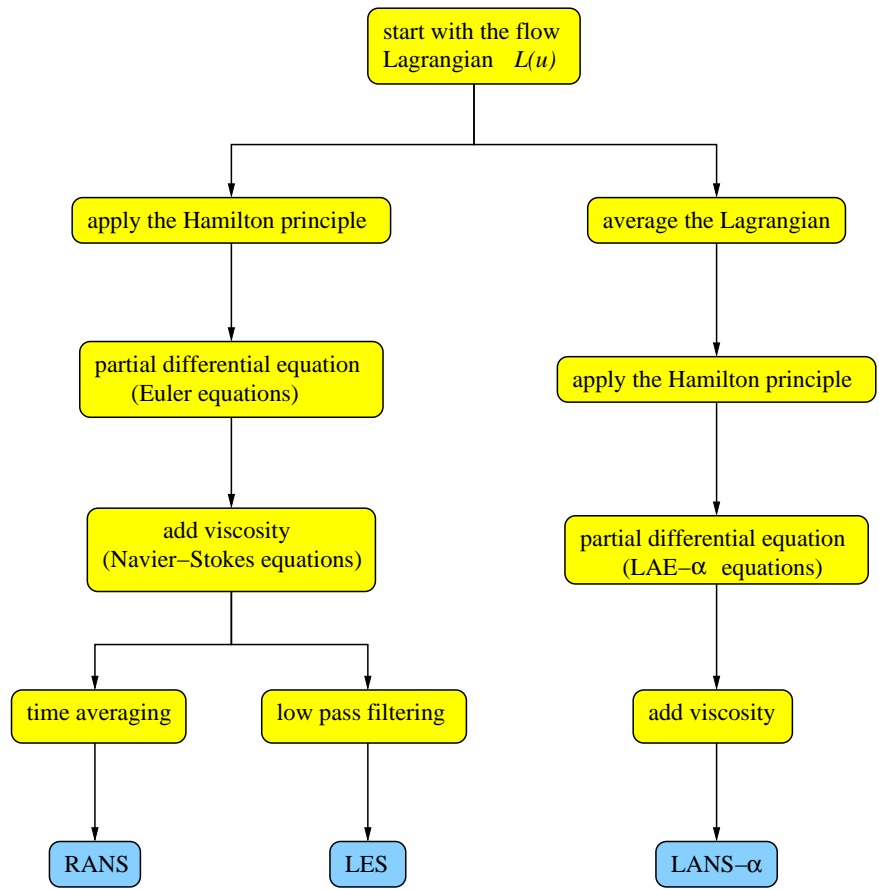}
\caption{Derivation of the averaged flow equations.}
\label{Fig1}
\end{center}
\end{figure}
More recently, Holm, Marsden and their coworkers \cite{Marsden:98b}
introduced a Lagrangian averaging technique for the mean motion of
ideal incompressible flows. Figure \ref{Fig1} contrasts the derivation
of LES, RANS, and the Lagrangian Averaged Navier-Stokes-$\alpha$
(LANS-$\alpha$) equations. Unlike the traditional averaging or
filtering approach used for both RANS and LES, where the Navier-Stokes
equations are averaged or spatially filtered, the Lagrangian averaging
approach is based on averaging at the level of the variational
principle. In the isotropic Lagrangian Averaged Euler-$\alpha$
(LAE-$\alpha$) equations, fluctuations smaller than a specified scale
$\alpha$ are averaged at the level of the flow maps
\cite{Mohseni:03c}. Mean fluid dynamics are derived by applying an
averaging procedure to the action principle of the Euler equations.
As shown in Figure \ref{Fig1}, both the Euler and the Navier-Stokes
equations can be derived in this manner (see Marsden \&
Ratiu~\cite{Marsden:98g} for a variational derivation of the Euler
equations).  The usual Reynolds Averaged Navier-Stokes (RANS) or LES
equations are then obtained through the subsequent application of
either a temporal or spatial average.  The critical difference with
the Lagrangian averaging procedure is that the Lagrangian (kinetic
energy minus potential energy) is averaged {\it prior to the
application of Hamilton principle and a closure assumption is applied
at this stage.}  This procedure results in either the Lagrangian
averaged Euler Equations (LAE-$\alpha$)\footnote{In this nomenclature,
$\alpha$ is used to denote the filtering scale (i.e. the simulation
faithfully represents motions on a scale larger than $\alpha$).}  or
the Lagrangian averaged Navier-Stokes Equations (LANS-$\alpha$),
depending on whether or not a random walk component is added in order
to produce a true molecular diffusion term.  Since the Hamilton
principle is applied after the Lagrangian averaging is performed, {\it
all the geometrical properties (\eg invariants) of the inviscid
dynamics are retained even in the presence of the model terms which
arise from the closure assumption} \cite{Marsden:98b, Marsden:98c,
Holm:99a}. For instance, LAE equations posses a Kelvin circulation
theorem. Thus it is potentially possible to model the transfer of
energy to the unresolved scales without an incorrect attenuation of
quantities such as resolved circulation.  This is an important
distinction for many engineering and geophysical flows where the
accurate prediction of circulation is highly desirable.

Numerical simulations by Chen \emph{et al} \cite{Chen:99b} and Mohseni
\emph{et al} \cite{Mohseni:03a} showed the capability of the
LANS-$\alpha$ equations in simulating isotropic homogenous turbulence.
However, most engineering and geophysical flows of interest are often
anisotropic. For example, due to rapid damping of turbulent
fluctuations in the vicinity of a wall, the application of the
isotropic LANS-$\alpha$ equations with a constant $\alpha$ is not
appropriate for long term calculations. In order to capture the
correct behavior in such systems the parameter $\alpha$ must be
spatially or/and temporally varied in the direction of anisotropy
\cite{chen:98a}, \ie wall normal direction.  There has been some
attempt (with limited success) in order to remedy this problem. A
successful {\it dynamic} LANS-$\alpha$ model is yet to be formulated
and tested. There are at least two approaches to anisotropy in the
LANS-$\alpha$ equations:
\begin{itemize}
\item[(i)] To derive a set of {\it anisotropic} LANS-$\alpha$
  equations. See alternative derivations in \cite{Holm:99a,
    Marsden:02b}.
\item[(ii)] Use the isotropic LANS-$\alpha$ equations, but with a variable
  $\alpha$ to compensate for the anisotropy.  
\end{itemize} 
At this point much more work must be done on the anisotropic
LANS-$\alpha$ equations before they can be applied to practical
problems. The second approach listed above is what will be explored in
this study.

This paper is organized as follows: The isotropic LANS-$\alpha$
equations and some of their main features are summarized in section
\ref{sec:LANS}.  A dynamic LANS-$\alpha$ approach is proposed in
section \ref{sec:DynLANS} where the variation in the parameter
$\alpha$ in the direction of anisotropy is determined in a
self-consistent way from the data contained in the simulation itself.
Our approach will be developed in the same spirit as the dynamic
modeling procedure for conventional LES~\cite{Germano:91a, Moin:95a,
Meneveau:97a, Lund:97a} which has achieved widespread use as very
effective means of estimating model parameters as a function of space
and time as the simulation progresses. The incompressible
Navier-Stokes equations are Helmholtz-filtered at the grid and a test
filter levels. A Germano type identity is derived by comparing the
filtered subgrid scale stress terms with those given in the
LANS-$\alpha$ equations. Considering a constant value of $\alpha$ and
averaging in the homogenous directions of the flow results in a
nonlinear equation for the parameter $\alpha$, which determines the
variation of $\alpha$ in the non-homogeneous directions or time. This
nonlinear equation is solved by an iterative technique. Consequently,
the parameter $\alpha$ is calculated during the simulation instead of
a fixed and pre-defined value.

Numerical techniques for simulating the dynamic LANS-$\alpha$ model in
this study are described in section \ref{sec:NumTechnique}. The
performance of the dynamic LANS-$\alpha$ model in simulating forced
and decaying isotropic homogeneous turbulent flows are considered in
section \ref{sec:Results}. In statistically equilibrated forced
turbulence, the parameter $\alpha$ should remain constant in time and
space. In decaying isotropic turbulence, the parameter $\alpha$ could
change in time as the integral scales of the turbulent flow changes.
In order to demonstrate the applicability of the dynamic LANS-$\alpha$
model of this study in anisotropic flows, {\it a priori} test of
turbulent channel flows are also performed in section
\ref{sec:Results}. Concluding results are presented in section
\ref{sec:Conclusions}.

\section{The Isotropic LANS-$\alpha$ Equations \label{sec:LANS}}
The incompressible isotropic LANS-$\alpha$ equations for the large
scale velocity $u$ are given by (see \cite{Marsden:98b} for a
derivation)
\begin{eqnarray}
  \dfrac{\p u}{\p t} + \( u\cdot \nabla \) u &=& - \nabla p +
    \dfrac{1}{Re} \Delta u + \nabla \cdot \tau(u),   
  \label{eq:LANS1} \\
  \nabla\cdot u & =& 0,
\end{eqnarray}
where $\tau(u)$ is the subgrid stress tensor defined as
\cite{Mohseni:03c}
\begin{equation}
  \tau(u) = -\alpha^2 (1-\alpha^2\Delta)^{-1} \left[ \nabla u\cdot
  \nabla u^T - \nabla u^T \cdot \nabla u + \nabla u \cdot \nabla u +
  \nabla u^T \cdot \nabla u^T \right].
  \label{eqn:Tau1}
\end{equation}
The subgrid scale stress $\tau(u)$ is in fact the momentum flux of the
large scales caused by the action of smaller, unresolved scales. Here
$\alpha$ is a constant length scale introduced during the averaging
process. Note that for vanishing parameter $\alpha$ the NS equations
will be recovered.

The LANS-$\alpha$ equations can be represented equivalently by
\begin{eqnarray}
      \frac{\partial v}{\partial t}
    + (u \cdot \nabla) v + v_j \nabla u_j = - \nabla P + \dfrac{1}{Re}
    \Delta v, \hspace{1cm} \textrm{where} \;\; v_{i} \;\; \textrm{is
    defined as} \hspace{1cm} v = u - \alpha^2 \Delta u. 
\label{eqn:euler}
\end{eqnarray}
The modified pressure $P$ in these equations 
is determined, as usual, from the incompressibility condition:
$\nabla\cdot u =0$ and $\nabla\cdot v =0$.

One interpretation for the equations (\ref{eq:LANS1}) is that they are
obtained by averaging the Euler equations in Lagrangian representation
over rapid fluctuations whose scale are of order $\alpha$. In this
respect, one can show that the Lagrangian averaged Euler equations can
be regarded as geodesic equations for the $H^1$ metric on the volume
preserving diffeomorphism group, as Arnold~\cite{Arnold:66a} did with
the $L_2$ metric for the Euler equations. Note that in calculating the
SGS stress $\tau(u)$ in equation (\ref{eqn:Tau1}) one needs to
calculate the inverse of the Helmholtz operator $(1-\alpha^2 \Delta)$,
which implies the need to solve a Poisson equation.  While efficient
numerical treatment of the Poisson equation, or its possible
elimination through rational approximation will be a focus of a future
publication, we note, in passing, that the inverse of the Helmholtz
operator can be expanded in $\alpha$ to higher orders of the Laplacian
operator as shown in below
\begin{equation}
  (1-\alpha^2 \Delta)^{-1}=1+\alpha^2\Delta+\alpha^4\Delta^2+\cdots. 
  \nonumber 
\end{equation}
Consequently, solving a Poission equation for inverting the Helmholtz
operator could be avoided.

It is interesting to note that the Lagrangian averaging technique
preserves the Hamiltonian structure of the governing equations in the
inviscid limit while the effects of small scales on the macroscopic
features of large scale are taken into account in a conservative
manner. The Hamiltonian and Lagrangian formulations of ideal fluids
are both basic and useful. These formulations are part of a more
general framework of {\it geometric mechanics}, which plays a vital
role in the development of new continuum models suited for
computation, as well as numerical algorithms that preserve structure
at the discrete level.  In recent years the geometric approach to
fluid mechanics has been quite successful.  Geometrical methods
provide a framework for the study of nonlinear stability
\cite{Marsden:84a},
variational integrators \cite{Marsden:00b, Marsden:01a},
statistical equilibrium theory \cite{Marsden:94b, Mohseni:01b}, and
many other interesting topics in fluid dynamics.  The Lagrangian
averaged Navier-Stokes-$\alpha$ uses ideas from geometric mechanics
and offers a theoretically and computationally attractive approach to
the turbulence closure problem.

\section{Derivation of a Dynamic LANS-$\alpha$ Model
  \label{sec:DynLANS}} 
The LANS-$\alpha$ equations for the large scale velocity $u$ are given
by equations (\ref{eq:LANS1}), where $\tau(u)$ is the subgrid stress
tensor defined in (\ref{eqn:Tau1}).  This set of equations for
$\alpha$ is similar to the grid filtered equation in the dynamic LES.
In analogy with the dynamic LES one can obtain an equation for the
filtering length scale, $\alpha$, by filtering the Navier-Stokes
equations
\begin{equation}
  \frac{\p u_i}{\p t}+u_j \frac{\p u_i}{\p x_j}=-\frac{\p p}{\p x_i}
  +\dfrac{1}{Re} \frac{\p^2 u_i}{\p x_j \p x_j},
  \label{eq_ns}
\end{equation}
with the Helmholtz related filters
\begin{eqnarray}
  \bar{u}& =&(1-\alpha^2\Delta)^{-1} u, \hspace{31mm}  
  \textrm{grid filter},  \label{eq:TestFilter1} \\
  \hat{\bar{u}}& =&(1-\widehat{\alpha}^2\Delta)^{-1}(1-\alpha^2
  \Delta)^{-1} u,
  \hspace{10mm} \textrm{test filter}, \label{eq:GridFilter1}
\end{eqnarray}
to obtain
\begin{equation}
\dfrac{\p \bar{u}_i}{\p t}+\dfrac{\p \bar{u}_i \bar{u}_j}{\p
  x_j}=-\dfrac{\p \bar{p}}{\p x_i}  +\dfrac{1}{Re} \dfrac{\p^2
  \bar{u}_i}{\p x_j \p x_j}-\dfrac{\p \tau_{ij}}{\p x_j},
\end{equation}
\begin{equation}
\dfrac{\p \hat{\bar{u}}_i}{\p t}+\dfrac{\p \hat{\bar{u}}_i
\hat{\bar{u}}_j}{\p x_j}=-\dfrac{\p \hat{\bar{p}}}{\p x_i}
+\dfrac{1}{Re} \dfrac{\p^2 \hat{\bar{u}}_i}{\p x_j \p x_j}-\dfrac{\p
  T_{ij}}{\p x_j},
\end{equation}
where
\begin{eqnarray}
  \tau_{ij} &=& \overline{u_i u_j}-\bar{u}_i \bar{u}_j, \nonumber \\
  T_{ij} &=& \widehat{\overline{u_iu_j}}-\hat{\bar{u}}_i
  \hat{\bar{u}}_j. \nonumber
\end{eqnarray}
Using an idea similar to Germano identity \cite{Germano:91a}, we
define
\begin{equation}
  L_{ij}=T_{ij}-\hat{\tau}_{ij}=\widehat{\bar{u}_i
  \bar{u}_j}-\hat{\bar{u}}_i \hat{\bar{u}}_j,
  \label{eq:Lij5}
\end{equation}
where the subgrid scale stresses under two filtering actions can be
modeled by the LANS-$\alpha$ subgrid term in equation
(\ref{eqn:Tau1}). Therefore,
\begin{equation}
\tau_{ij}=\alpha^2(1-\alpha^2 \Delta)^{-1} M_{ij},
\label{eq:Tgauij5}
\end{equation}
\begin{equation}
  T_{ij}=\hat{\alpha}^2(1-\hat{\alpha}^2 \Delta)^{-1} N_{ij},
\label{eq:Tij5}
\end{equation}
where
\begin{eqnarray}
  M_{ij} &=& \dfrac{\p\bar{u}_i}{\p x_k}\dfrac{\p\bar{u}_j}{\p x_k}
  -\dfrac{\p\bar{u}_k}{\p x_i}\dfrac{\p\bar{u}_k}{\p
  x_j}+\dfrac{\p\bar{u}_i}{\p x_k}\dfrac{\p\bar{u}_k}{\p x_j}
+\dfrac{\p\bar{u}_j}{\p x_k}\dfrac{\p\bar{u}_k}{\p x_i},
  \nonumber \\
  N_{ij} &=& \dfrac{\p\hat{\bar{u}}_i}{\p x_k}
  \dfrac{\p\hat{\bar{u}}_j}{\p x_k} - \dfrac{\p\hat{\bar{u}}_k}{\p
  x_i} \dfrac{\p\hat{\bar{u}}_k}{\p x_j}+ \dfrac{\p\hat{\bar{u}}_i}{\p
  x_k} \dfrac{\p\hat{\bar{u}}_k}{\p x_j} + \dfrac{\p\hat{\bar{u}}_j}{\p
  x_k} \dfrac{\p\hat{\bar{u}}_k}{\p x_i}.
 \nonumber 
\end{eqnarray}
Combining equations (\ref{eq:Lij5})-(\ref{eq:Tij5}), one obtains
\begin{equation}
  L_{ij}=\beta^2\alpha^2(1-\beta^2\alpha^2 \Delta)^{-1}N_{ij}-
  \alpha^2(1-\beta^2\alpha^2 \Delta)^{-1}(1-\alpha^2 \Delta)^{-1}M_{ij},
\end{equation}
or
\begin{equation}
  L_{ij}=\alpha^2(\beta^2\hat{N}_{ij}-\hat{\bar{M}}_{ij}),
\end{equation}
where $\beta=\hat{\alpha}/\alpha$. Multiplying both sides of the above
equation by $S_{ij}$, to yield
\begin{equation}
  L_{ij}S_{ij}=\alpha^2(\beta^2\hat{N}_{ij}-\hat{\bar{M}}_{ij})S_{ij}.
\end{equation} 
Taking spatial averaging of both sides of the above equation in homogenous
directions, one obtains
\begin{equation}
  \alpha^2=\dfrac{\langle L_{ij}S_{ij} \rangle}
  {\langle (\beta^2\hat{N}_{ij}-\hat{\bar{M}}_{ij})S_{ij}\rangle},
  \label{eq:Alpha0}
\end{equation}
where
\begin{equation}
  S_{ij}=\dfrac{1}{2} \( \dfrac{\p \bar{u}_i}{\p x_j}+ \dfrac{\p
  \bar{u}_j}{\p x_i} \).
  \nonumber
\end{equation}
The denominator in equation (\ref{eq:Alpha0}) could approach zero,
where it creates a singularity. In dynamic LES, Lilly \cite{Lilly:92a}
used a least square approach to eliminate the singularity in Germano's
model. By a similar least square approach a nonlinear equation for
$\alpha$ could be found as
\begin{equation}
\alpha^2=F(\alpha)=\dfrac{\langle
  L_{ij}(\beta^2\hat{N}_{ij}-\hat{\bar{M}}_{ij}) \rangle} {\langle
  (\beta^2\hat{N}_{ij}-\hat{\bar{M}}_{ij})(\beta^2\hat{N}_{ij}-
  \hat{\bar{M}}_{ij})\rangle}, 
\label{eq:Alpha1}
\end{equation}
which does not have the singularity problem as in equation
(\ref{eq:Alpha0}). This is a nonlinear equation for $\alpha$. All the
quantities in equation (\ref{eq:Alpha1}) can be calculated during a
LANS-$\alpha$ simulation. Therefore, equation (\ref{eq:Alpha1})
provides a nonlinear equation for dynamically calculating the value of
$\alpha$  during the simulation. 

At this point the potential values for the free parameter $\beta$ are
required. Writing the grid and test filters in equations
(\ref{eq:TestFilter1}) and (\ref{eq:GridFilter1}) in the Fourier
space, one obtains
\begin{equation}
\check{\bar{u}}=\dfrac{\check{u}}{1+\alpha^2 k^2},
\end{equation}
and
\begin{equation}
\check{\hat{\bar{u}}}=\dfrac{\check{u}}{(1+\beta^2 \alpha^2
  k^2)(1+\alpha^2 k^2)} \approx \dfrac{\check{u}}{1+(\beta^2+1)
  \alpha^2 k^2} =\dfrac{\check{u}}{1+\tilde{\alpha}^2 k^2} \;\;\;\;
  \text{as} \;\;\;\; k \rightarrow \infty,
\end{equation}
\begin{figure}
\begin{center}
\includegraphics[width=.5\linewidth]{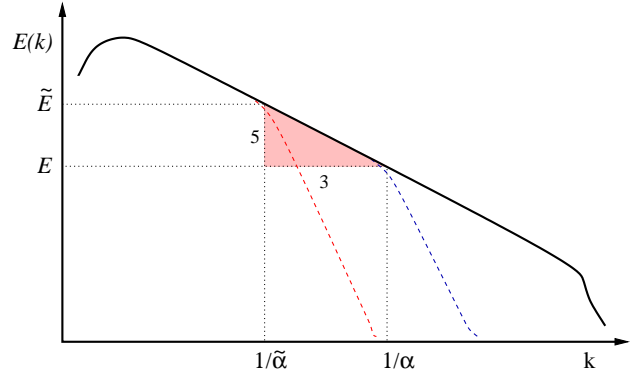}
\end{center}
\caption{The positions of grid and test filter scales on the turbulent
  kinetic energy spectrum. \label{energy_spectra}}
\end{figure}
where $\check{(\cdot)}$ stands for variables in the Fourier space, $k$
is the wavenumber, and $\tilde{\alpha}$ corresponds to filter scale
for the test filter.  Since $\tilde{\alpha}=\sqrt{1+\beta^2} \alpha
\geq \alpha$, one can realize that as long as $\beta > 0$, the test
filter have a larger filter scale than the grid filter. Figure
\ref{energy_spectra} shows the relative positions of the grid filter
scale $\alpha$ and the test filter scale $\tilde{\alpha}$ on a
schematic of the energy spectrum for a high Reynolds number flow. In
order to accurately model the subgrid scale stress, both the grid
filter and the test filter scales must be located in the inertial
sub-range of the energy spectrum. It should be pointed out that the
iterative calculation required in equation (\ref{eq:Alpha1}) does not
require new flow field calculations, and the iteration at each time
step is carried out using the existing flow field at the same time
step.  Similar to the dynamic LES model, the present dynamic
LANS-$\alpha$ model has a free parameter $\beta$, which is related to
the characteristic length scale of the grid and test filters.

The dynamic $\alpha$ model given in equation (\ref{eq:Alpha1}) is
designed to capture the length scale variations in space and time.
Aside from the isotropic homogenous turbulent flows, it is well suited
for anisotropic flows such as wall bounded turbulence or mixing flow
turbulence, where the turbulence length scales could change in space
or in time. In cases where there are directions of homogeneity, such
as the streamwise and spanwise direction in a channel flow, one can
average the parameter $\alpha$ over the homogeneous directions. In a
more general situation, we expect to replace the plane average, used in
the channel flow, by an appropriate local spatial or time averaging
scheme. For isotropic homogenous turbulence, $\alpha$ is regarded
as a constant in space and changes only in time.  

\section{Numerical Method \label{sec:NumTechnique}}
The dynamic procedure in this study is initially tested for forced and
decaying isotropic turbulence where the parameter $\alpha$ is constant
over the computational domain, but can vary in time. Furthermore, {\it
a priori} test of the dynamic LANS-$\alpha$ procedure in a turbulent
channel flow is investigated. In this section the numerical technique
for solving the governing equations are summarized.

\paragraph{Isotropic homogeneous turbulence.} The computations are
performed in a periodic cubic box of side $2\pi$. A standard parallel
pesudospectral scheme with periodic boundary conditions are
employed. The spatial derivatives are calculated in the Fourier
domain, while the nolinear convective terms are computed in the
physical space. A fourth order Runge-Kutta scheme is implemented to
advance the flow field in time. The two third rule is used in order to
eliminate the aliasing errors. Therefore, the upper one third of the
wave modes are discarded at each stage of the Runge-Kutta scheme.  The
initial velocity field for each case was divergence free and
constructed to generate an energy spectrum of the form
\begin{equation}
  E(k)\sim k^4 exp[-2(k/k_p)^2].
  \nonumber
\end{equation}
The value of $k_p$ corresponds to the peak in the energy spectrum.
The initial pressure fluctuations were obtained from the solution of a
Possion equation.

\paragraph{Turbulent flow in a channel.} DNS data from del \'Alamo and
J. Jim\'enez \cite{Jimenez:03a} are employed for the {\it a priori}
test. The computational domain in this case, normalized based on the
half channel height, is spanned $8\pi$ in the streamwise and $4\pi$ in the
spanwise directions. The spatial derivatives are calculated by the
pesudospectral method in streamwise and spanwise directions and by the
Chebychev-tau technique in the wall normal direction. Similar
computational techniques have successfully been used for the DNS
of channel flows by Kim \emph{et al} \cite{Moin:87b} and Moser
\emph{et al} \cite{Moser:99a}.  Grid and test filters of Helmholtz
types are applied in both streamwise and spanwise directions, while no
explicit filters are applied in the wall normal direction.  $\alpha$
is assumed to be constant in the homogenous directions, \ie the
streamwise and spanwise directions, in order to solve the nonlinear
equations (\ref{eq:Alpha0}) or (\ref{eq:Alpha1}). These equations are
solved by an iterative technique. Since both the mean flow and the
flow perturbations vanish at the wall, singular behavior might occur
in these equations.  This can be easily fixed by starting the {\it a
priori} test a few grid points away from the wall. In actual
simulation of the dynamic LANS-$\alpha$ equations, one can explicitly
put $\alpha$ to zero below in the immediate vicinity of a wall when
the value of $\alpha$ drops below a threshold.
The converged $\alpha$ values at each point is used as an initial
value for the iteration at the next grid layer.

\section{Results and Discussions \label{sec:Results}}
Capabilities of the dynamic LANS-$\alpha$ model of the previous
sections are examined in both isotropic and anisotropic turbulent
flows. In isotropic homogeneous turbulence the parameter $\alpha$ is
constant in space but allowed to vary in time. Results of the dynamic
model is compared with the isotropic LANS-$\alpha$ simulations with a
constant $\alpha$ and with the DNS data.

\begin{figure}[hbt]
\begin{center}
  \begin{minipage}{0.48\linewidth} \begin{center}
  \epsfig{file=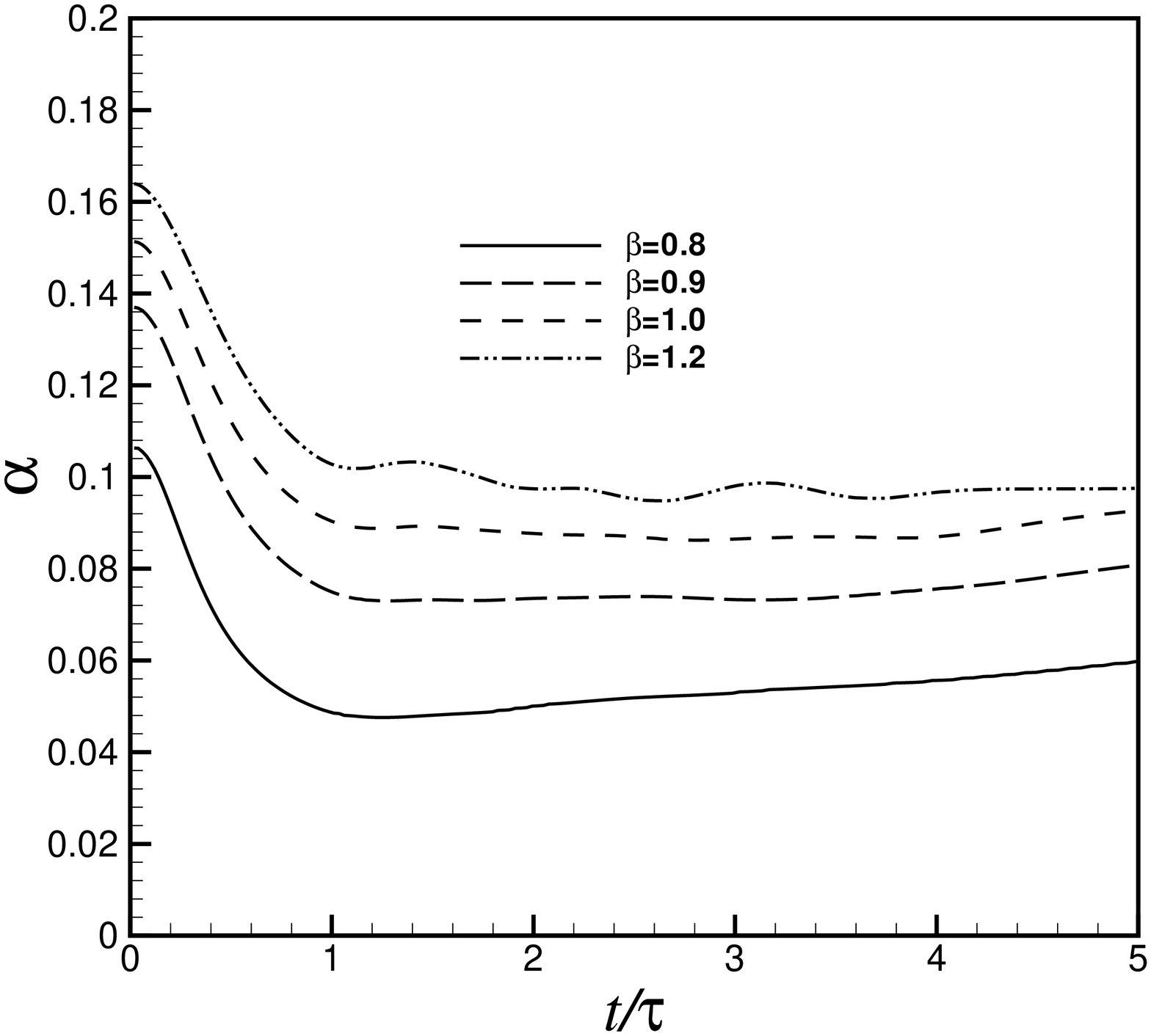}, width=0.99\linewidth}
  \end{center} \end{minipage}
  \begin{minipage}{0.48\linewidth} \begin{center}
  \epsfig{file=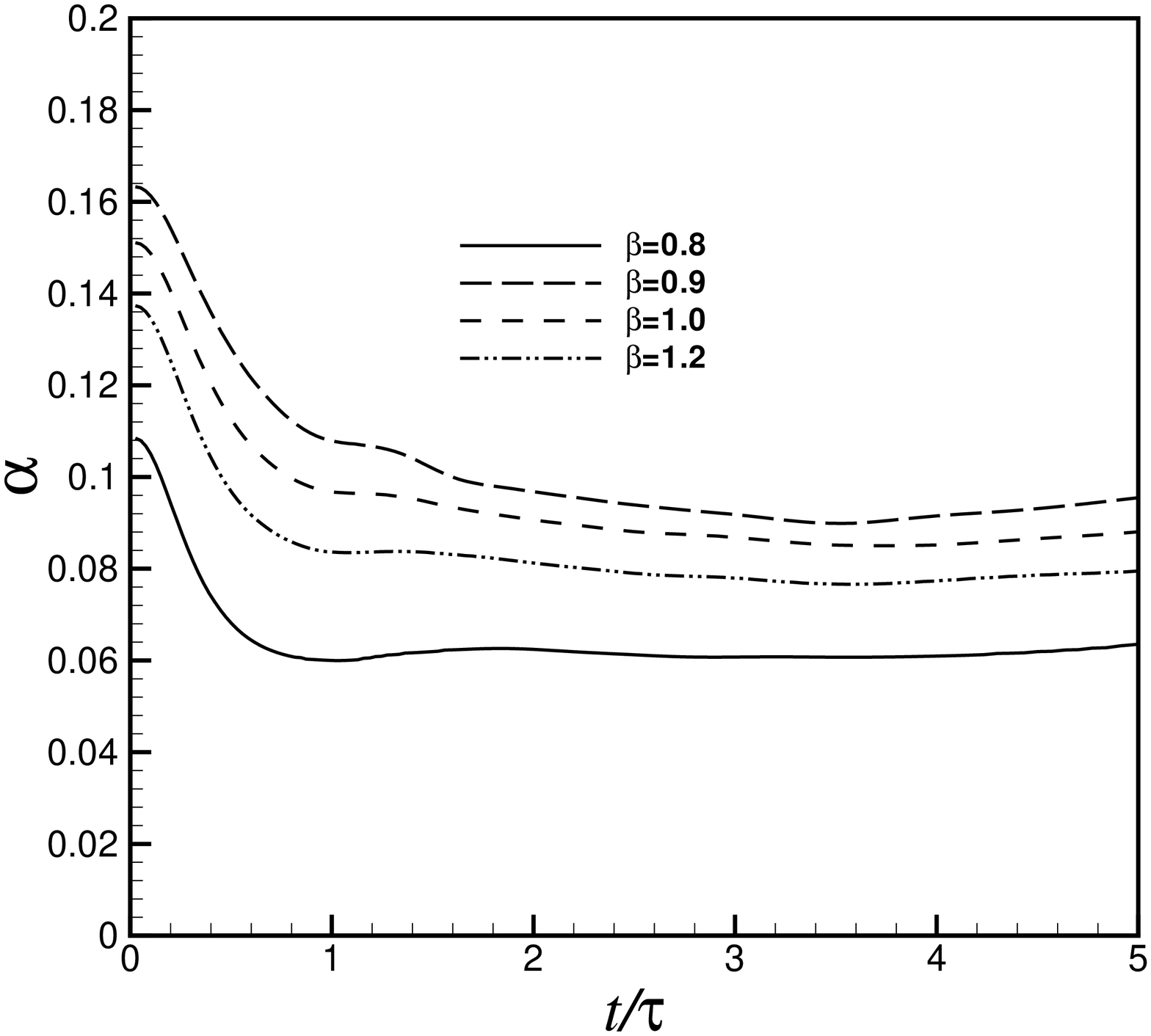}, width=0.99\linewidth} 
  \end{center}\end{minipage}\\
\begin{minipage}{0.48\linewidth} \begin{center} (a) \end{center} \end{minipage}
\begin{minipage}{0.48\linewidth} \begin{center} (b) \end{center} \end{minipage}
\caption{Evolution of $\alpha$ for different $\beta$ for a decaying
 isotropic turbulence at $Re_\lambda=72$ and $\tau=0.9$. Grid
 resolution (a) $48^3$, (b) $32^3$.}
\label{decay_alp}
\end{center}
\end{figure}

\paragraph{Decaying isotropic homogenous turbulence simulations.} DNS
of a decaying isotropic homogenous turbulence with initial Taylor
Reynolds number of $Re_\lambda=72$ (corresponding to a computational
Reynolds number $Re=300$) is performed to be used as a test case. The
initial energy spectrum is peaked at $k_p=4$. The isotropic
LANS-$\alpha$ and the dynamic LANS-$\alpha$ simulations are calculated
for both $64^3$ (corresponds to $48^3$ after dealiasing) and $48^3$
(corresponds to $32^3$ after dealiasing) resolutions, and direct
numerical simulations are performed for $128^3$ (corresponds to $85^3$ 
after dealiasing). The eddy turn over
time for this case is found to be $\tau=0.9$.  Figure \ref{decay_alp}
shows the time evolution of $\alpha$ for $\beta=0.8, 0.9, 1,$ and 1.2.
The values of $\alpha$ experience a sharp decrease from its initial
value during the first eddy turn over time. However, it quickly settle
down toward a much slower varying value.  Slight changes in $\alpha$
value after the first eddy turn over time could be traced back to
flattening of the energy spectrum as the turbulence decays.

\begin{figure}[hbt]
\begin{center}
\begin{minipage}{0.48\linewidth} \begin{center}
  \epsfig{file=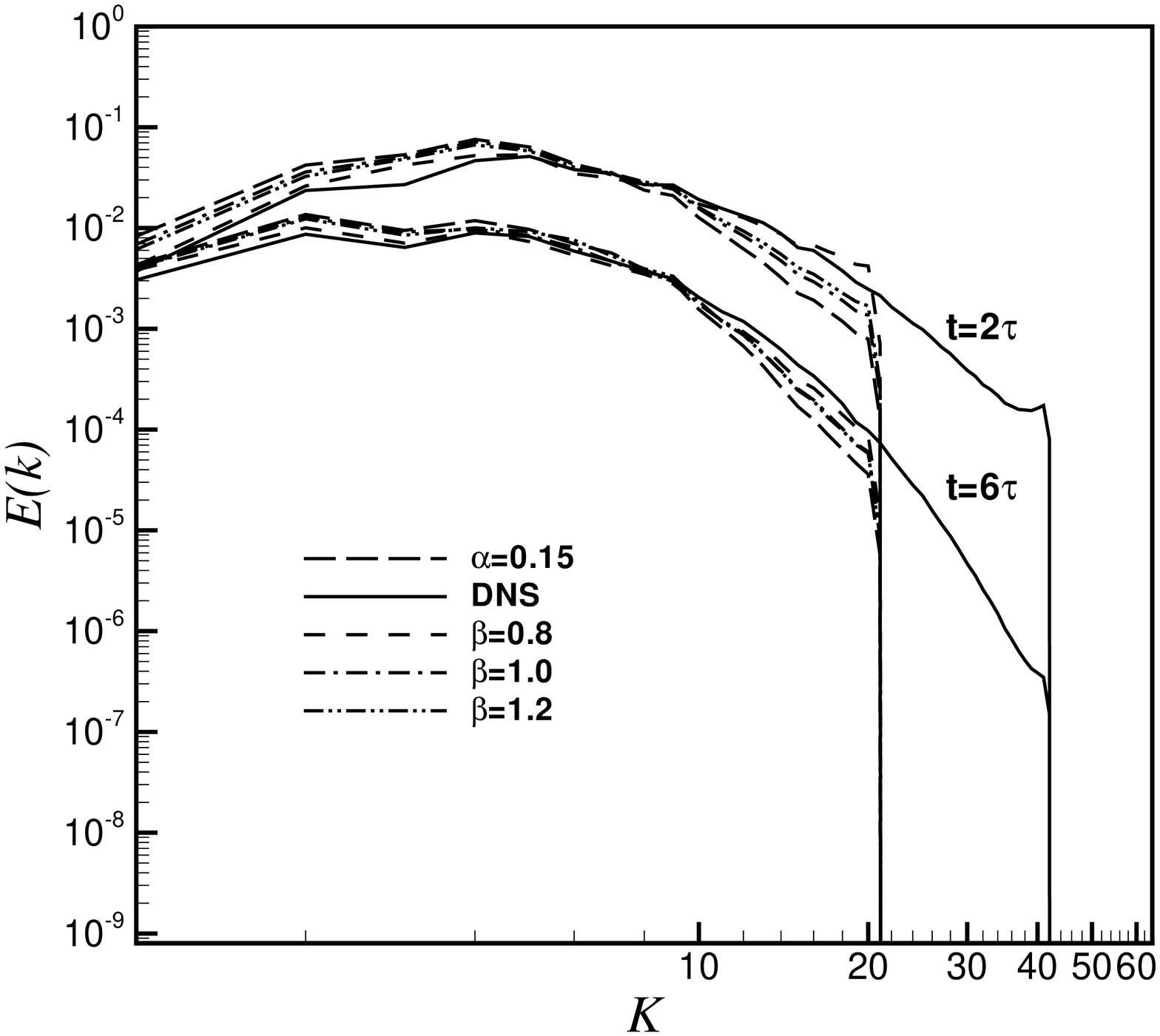}, width=0.99\linewidth} 
\end{center} \end{minipage}
\begin{minipage}{0.48\linewidth} \begin{center}
  \epsfig{file=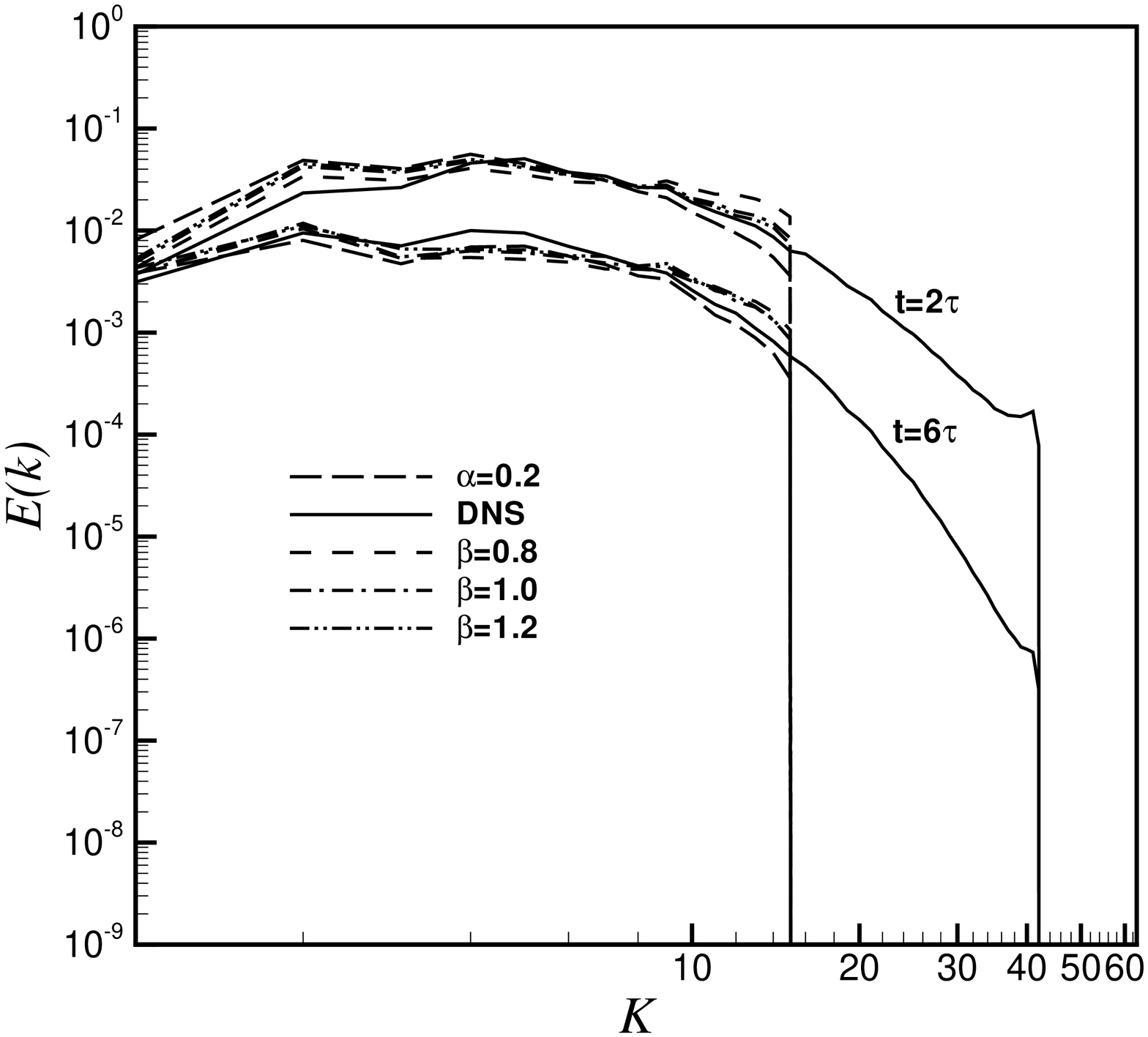}, width=0.99\linewidth} 
\end{center}\end{minipage}\\
\begin{minipage}{0.48\linewidth} \begin{center} (a) \end{center} \end{minipage}
\begin{minipage}{0.48\linewidth} \begin{center} (b) \end{center} \end{minipage}
\caption{Energy spectra of the DNS, dynamic LANS-$\alpha$, and
LANS-$\alpha$ with fixed $\alpha$ simulations of a decaying isotropic
turbulence at $Re_\lambda=72$ and $\tau=0.9$. Grid resolution (a)
$48^3$, (b) $32^3$.}
\label{decay_energy_spectra}
\end{center}
\end{figure}

\begin{figure}[hbt]
\begin{center}
  \begin{minipage}{0.48\linewidth} \begin{center}
    \epsfig{file=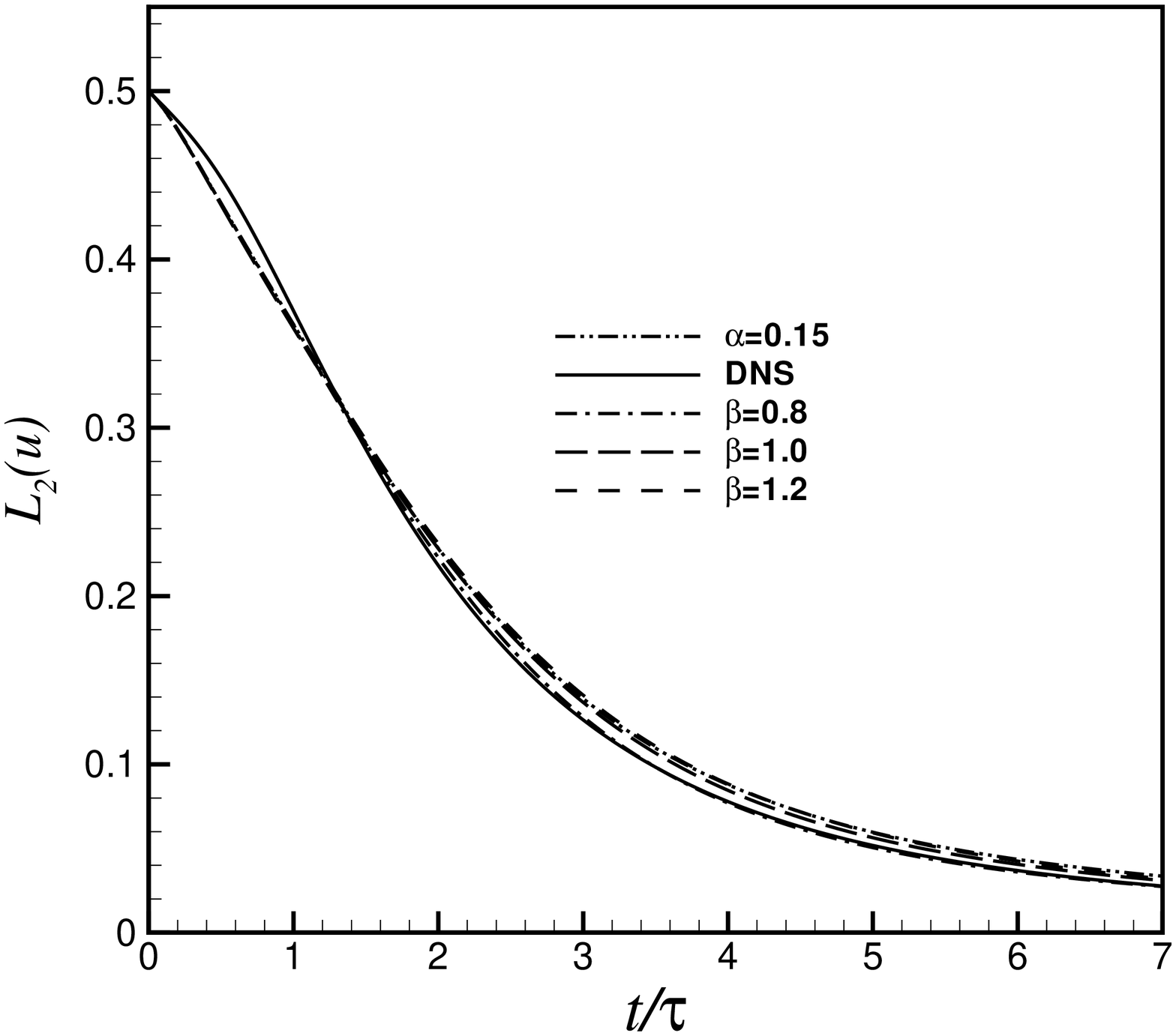}, width=0.99\linewidth} 
  \end{center} \end{minipage}
  \begin{minipage}{0.48\linewidth} \begin{center}
    \epsfig{file=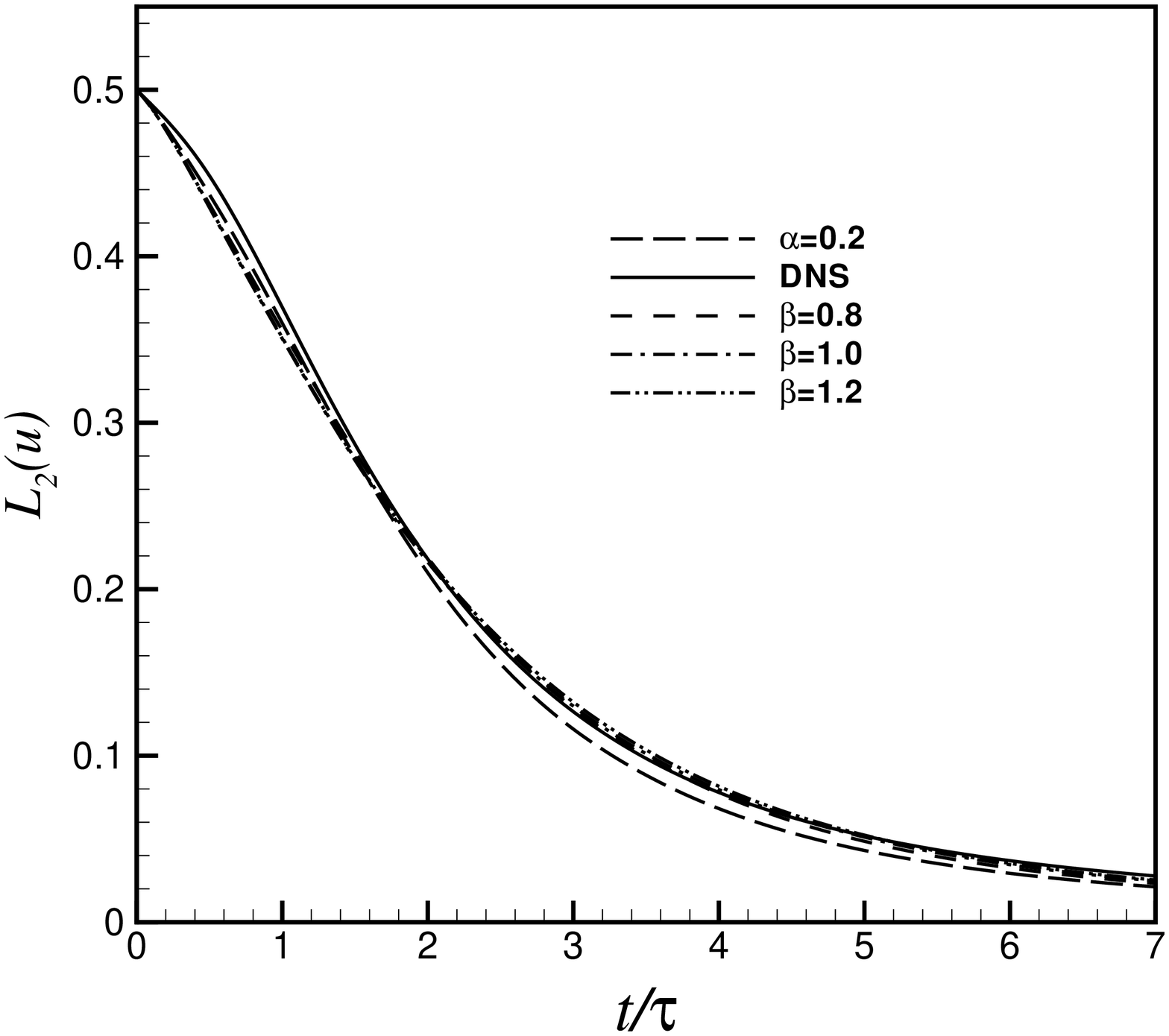}, width=0.99\linewidth} 
  \end{center}\end{minipage}\\
\begin{minipage}{0.48\linewidth} \begin{center} (a) \end{center} \end{minipage}
\begin{minipage}{0.48\linewidth} \begin{center} (b) \end{center} \end{minipage}
\caption{Total kinetic energy decay of the DNS, dynamic LANS-$\alpha$,
and LANS-$\alpha$ with fixed $\alpha$ simulations of a decaying
isotropic turbulence at $Re_\lambda=72$ and $\tau=0.9$. Grid
resolution (a) $48^3$, (b) $32^3$.}
\label{total_energy_decay}
\end{center}
\end{figure}

The energy spectra at two different times are shown in Figure
\ref{decay_energy_spectra}, and the total kinetic energy decay are
shown in Figure \ref{total_energy_decay}. While a slight dependency on
the value of $\beta$ is observed, in general, the energy spectrum at
various times and the total kinetic energy decay are captured nicely.
Mohseni \emph{et al} \cite{Mohseni:03a} demonstrated that in order to
accurately simulate a turbulent flow with the LANS-$\alpha$
equations, the value of $\alpha$ should be somewhere, perhaps one
decade lower than the peak of the energy spectra toward the grid
resolution. Careful considerations of Figures \ref{decay_alp} and 
\ref{decay_energy_spectra} reveal that the dynamic
LANS-$\alpha$ model of this study satisfies this criteria for all
$\beta$ values. In general, one expects that the value of $\alpha$ to
be in the inertial range of the energy spectra in order to correctly
capture the dynamics of the large scales.   As illustrated in Figures
\ref{decay_energy_spectra} and \ref{total_energy_decay}, it is evident
that the dynamic LANS-$\alpha$ model provides a better estimate of the
total kinetic energy decay and the energy spectra over similar
simulations with fixed $\alpha$ calculations.

\paragraph{Forced isotropic homogenous turbulence simulations.} 
Forced isotropic turbulence is one of the most idealized and
extensively simulated turbulent flows. The numerical forcing of a
turbulent flow is usually referred to the artificial addition of
energy at the large scales in a numerical simulation.  Statistical
equilibrium is signified by the balance between the input of kinetic
energy through the forcing and its output through the viscous
dissipation. In this study, we adopted a forcing method used in Chen
\emph{et al} \cite{Chen:99b} and Mosheni \emph{et al}
\cite{Mohseni:03a} where the wave modes in a spherical shell $|K|=k_0$
of certain width are forced in such a way that the forcing spectrum
follows the Kolmogorov $-5/3$ scaling law, that is
\begin{equation}
  \check{f}_i=\frac{\delta_0}{N}\frac{\check{u}_i}{\sqrt{\check{u}_k
    \check{u}^*_k}} k^{-5/3}. 
\end{equation}
Here $\check{f}_i$ and $\check{u}_i$ are Fourier transforms of the
forcing vector $f_i$ and velocity $u_i$, $N$ is the number of forced
wave modes, and $\delta$ controls the injection rate of energy at the
large scales. This particular forcing technique enforces the energy
cascade in the inertial range starting from the first wave mode. In
this simulations we choose $k_0=2$ and $\delta_0=0.1$. The initial
Taylor Reynolds number is $Re_\lambda=415$ and the initial energy
spectrum is peaked at $k_p=1$, while the eddy turn over time is found
to be $\tau=3.8$. The grid resolution for simulations using the
dynamic LANS-$\alpha$ equations and the LANS-$\alpha$ equations with
fixed $\alpha$ is $64^3$, while the DNS data is performed at a grid
resolution of $128^3$ before dealiasing.

\begin{figure}[hbt]
\begin{center}
\epsfig{file=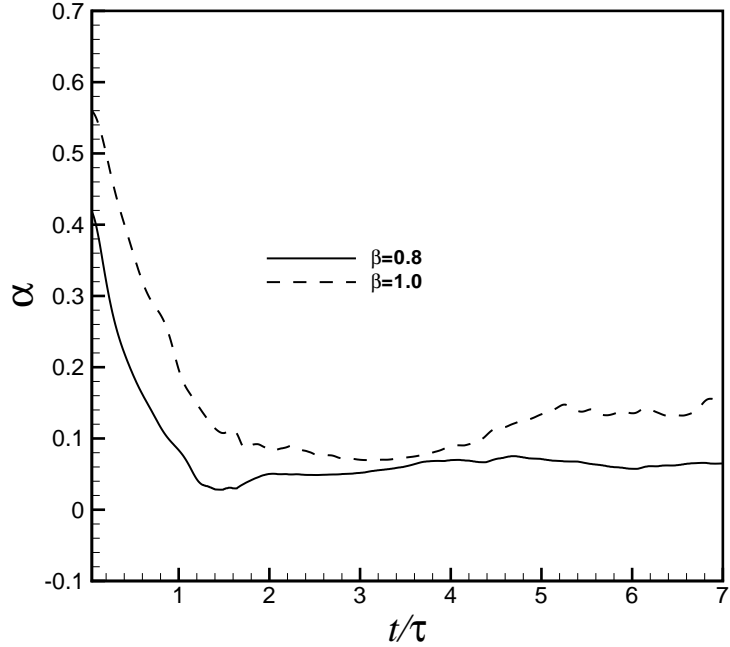}, width=0.6\linewidth}
\caption{Evolution of $\alpha$ for different $\beta$ in the forced
  turbulence case with $Re_\lambda=415$ and $\tau=3.8$.}
\label{forced_alp}
\end{center}
\end{figure}

Figure \ref{forced_alp} shows the evolution of $\alpha$ for
$\beta=0.8$ and 1.0.  Similar to the decaying turbulence, a sharp
decrease in the value of $\alpha$ is observed over the first eddy turn
over time, where the values of $\alpha$ settles down toward a constant
value. This corresponds to an statistically equilibrated state. As
expected, the final value of $\alpha$ is in the inertial range of the
energy spectrum.

\begin{figure}[hbt]
 \begin{center}
  \begin{minipage}{0.6\linewidth} \begin{center}
    \epsfig{file=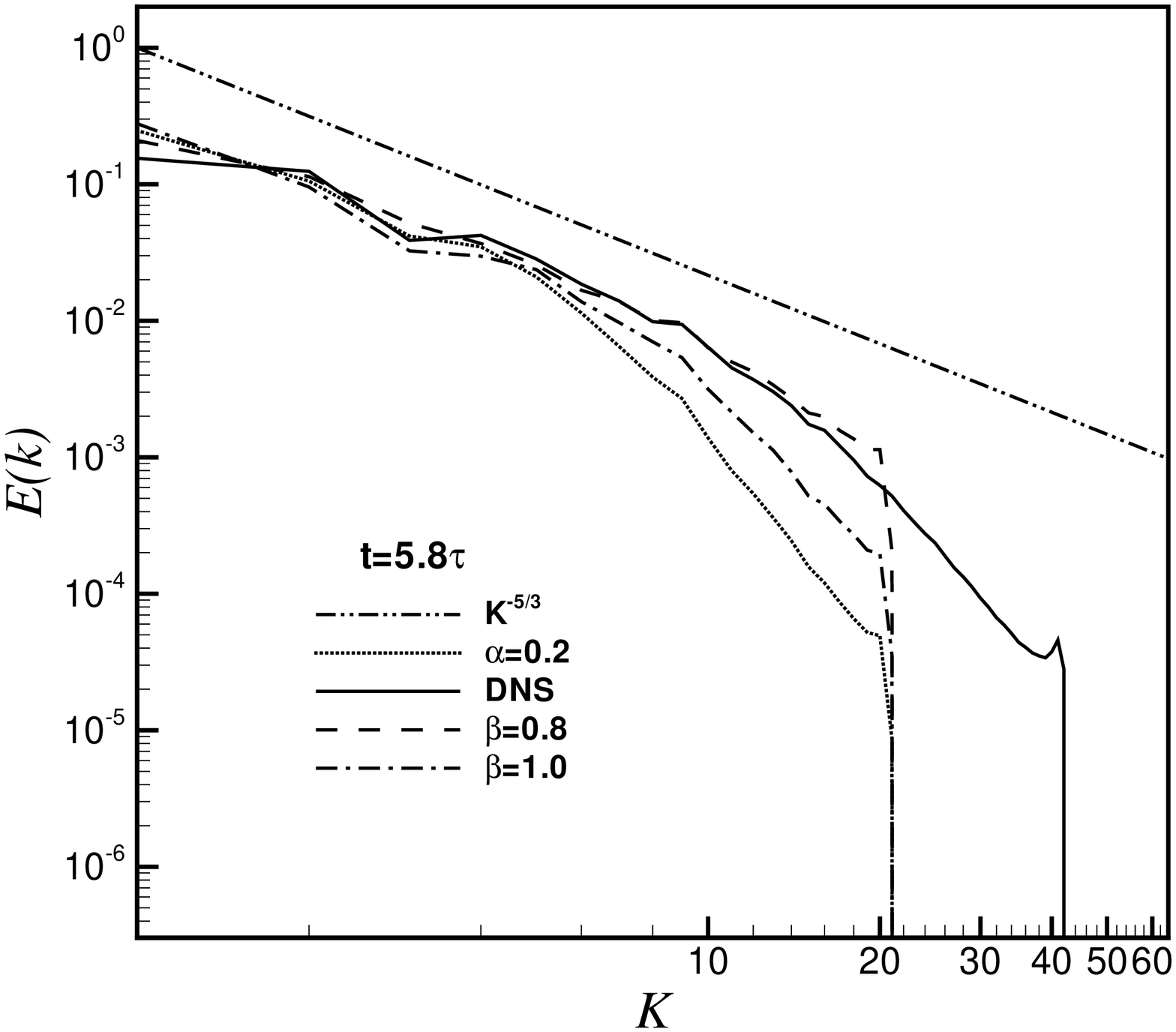}, width=0.99\linewidth} 
  \end{center}\end{minipage}
\caption{Energy spectra at $t=5.8 \tau$ for the DNS, dynamic
LANS-$\alpha$ and LANS-$\alpha$ with fixed $\alpha$ simulations of a forced isotropic turbulence with
$Re_\lambda=415$ and $\tau=3.8$.}
\label{forced_energy_spectra}
\end{center}
\end{figure}

Figure \ref{forced_energy_spectra}
shows the energy spectrum at $t=5.8 \tau$ for $\beta=0.8$ and 1.0.  An
inertial subrange with $\sim k^{-5/3}$ energy spectrum is evident in
the dynamic LANS-$\alpha$ simulations. The results of the dynamic
LANS-$\alpha$ simulations are compared with the DNS and the
LANS-$\alpha$ simulations with $\alpha=0.2$.  The energy spectra of
the dynamic LANS-$\alpha$ simulations for $\beta=0.8$ and 1.0 show a
better agreement with the DNS data than the energy spectra for a
LANS-$\alpha$ simulation with a constant $\alpha$.

\begin{figure}[hbt]
\begin{center}
  \begin{minipage}{0.48\linewidth} \begin{center}
    \epsfig{file=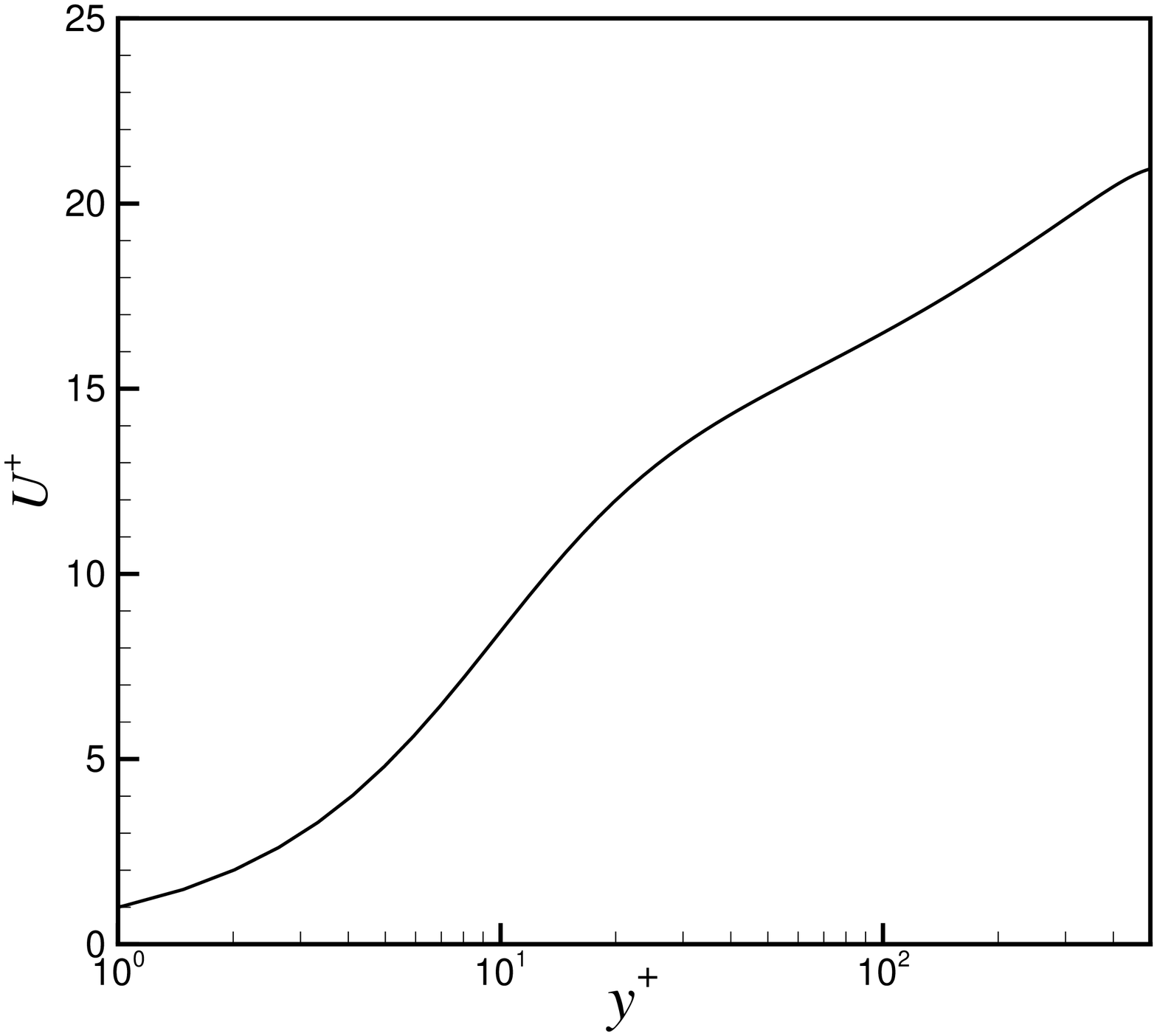}, width=0.98\linewidth}
  \end{center} \end{minipage}
  \begin{minipage}{0.48\linewidth} \begin{center}
    \epsfig{file=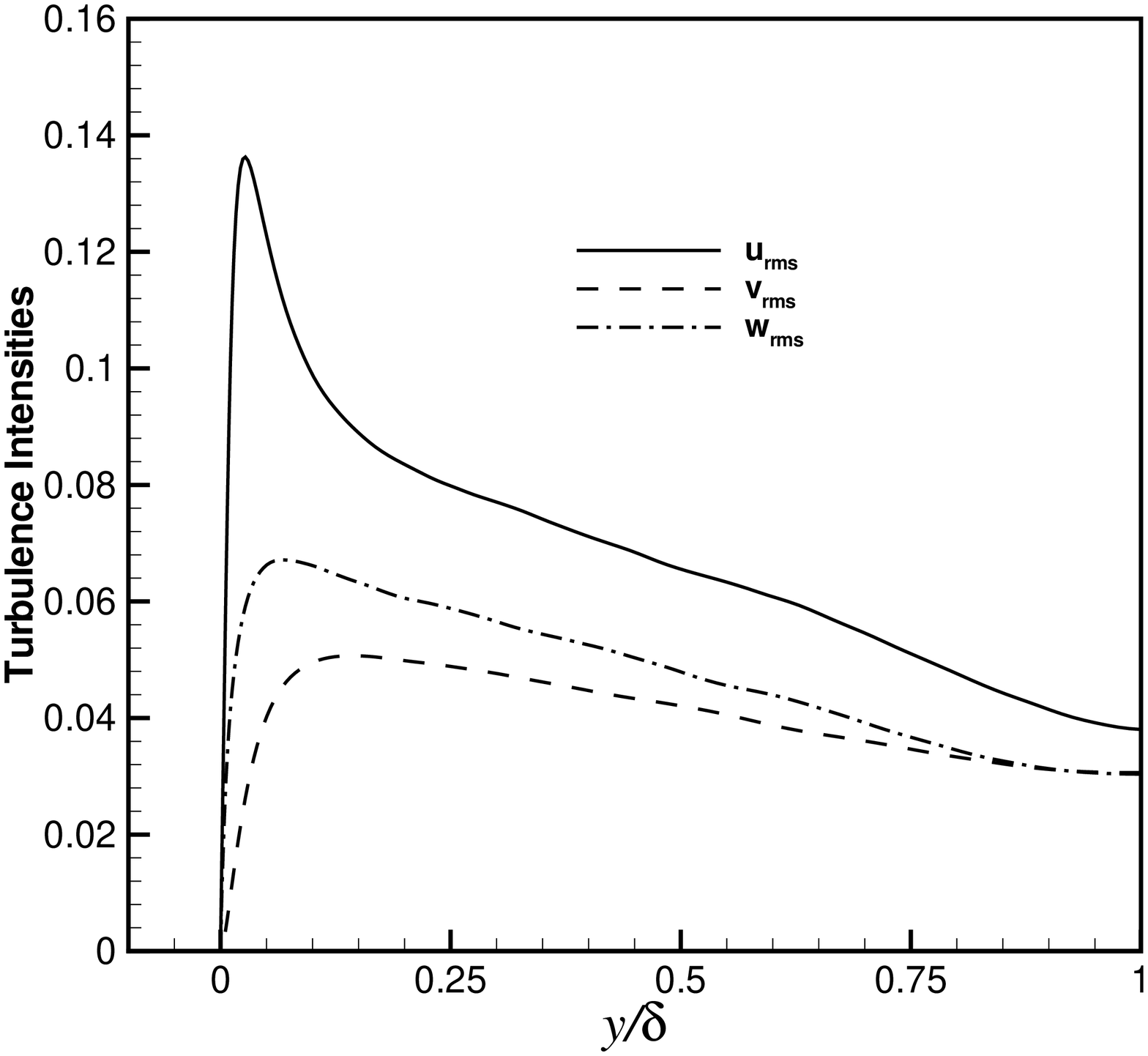}, width=0.98\linewidth}
  \end{center}\end{minipage}\\
\begin{minipage}{0.48\linewidth} \begin{center} (a) \end{center} \end{minipage}
\begin{minipage}{0.48\linewidth} \begin{center} (b) \end{center} \end{minipage}
\caption{DNS results of a turbulent channel flow at $Re_\tau=550$ from
  del \'Alamo and Jim\'enez \cite{Jimenez:03a}. (a) The mean velocity
  profile, (b) Root-mean-square velocity fluctuations in global
  coordinates.}
\label{umean_yplus_turbulence_intensities}
\end{center}
\end{figure}

\paragraph{{\it A priori} test of a turbulent channel flow.}
{\it A priori} test of the dynamic LANS-$\alpha$ model is carried out
in order to determine the accuracy of the model in predicting the SGS
stresses and the energy dissipation rates in a wall bounded flow. The
tests are performed on a DNS data of del \'Alamo and Jim\'enez
\cite{Jimenez:03a} for a turbulent channel flow. The turbulence
Reynolds number, based on the wall friction velocity, is $Re_\tau=550$
and the computational grid is $1536\times 257 \times 1536$ in the
streamwise, wall normal, and spanwise directions, respectively. After
dealiasing the physically relevant part of the computational domain
reduces to $1024\times 257 \times 1023$.  The mean velocity profile,
non-dimensionalized by the wall-shear velocity, is depicted in Figure
\ref{umean_yplus_turbulence_intensities}(a), where a log layer from
$y^+ \approx 80$ to 220 is observed. Figure
\ref{umean_yplus_turbulence_intensities}(b) shows the turbulence
intensity profiles from the wall to the middle of the channel in
global coordinate which is normalized by half channel height $\delta$. 
Maximum turbulence intensities in all directions
are located in the wall layer.

\begin{figure}[hbt]
\begin{center}
  \begin{minipage}{0.48\linewidth} \begin{center}
    \epsfig{file=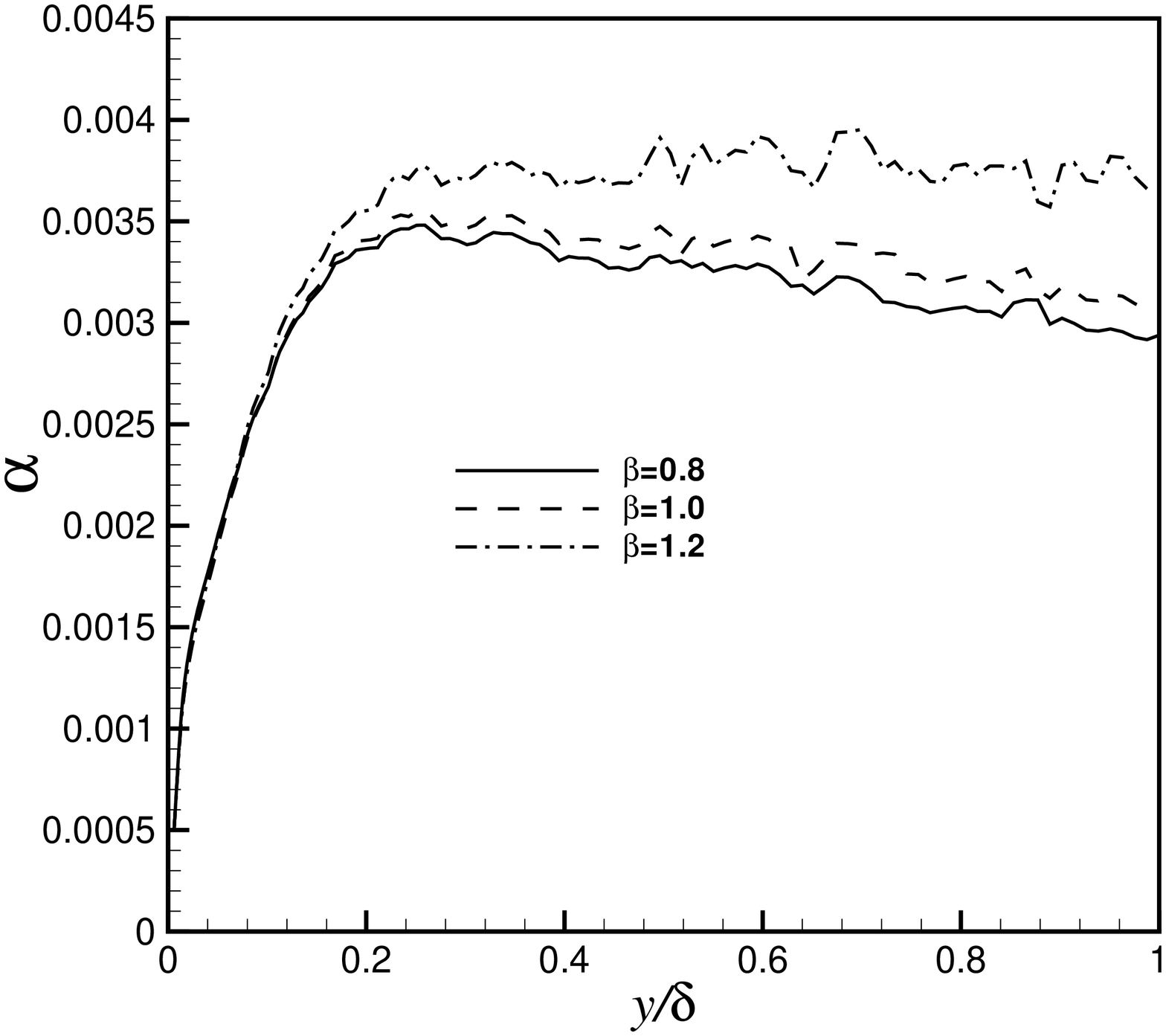}, width=0.98\linewidth}
  \end{center} \end{minipage}
  \begin{minipage}{0.48\linewidth} \begin{center}
    \epsfig{file=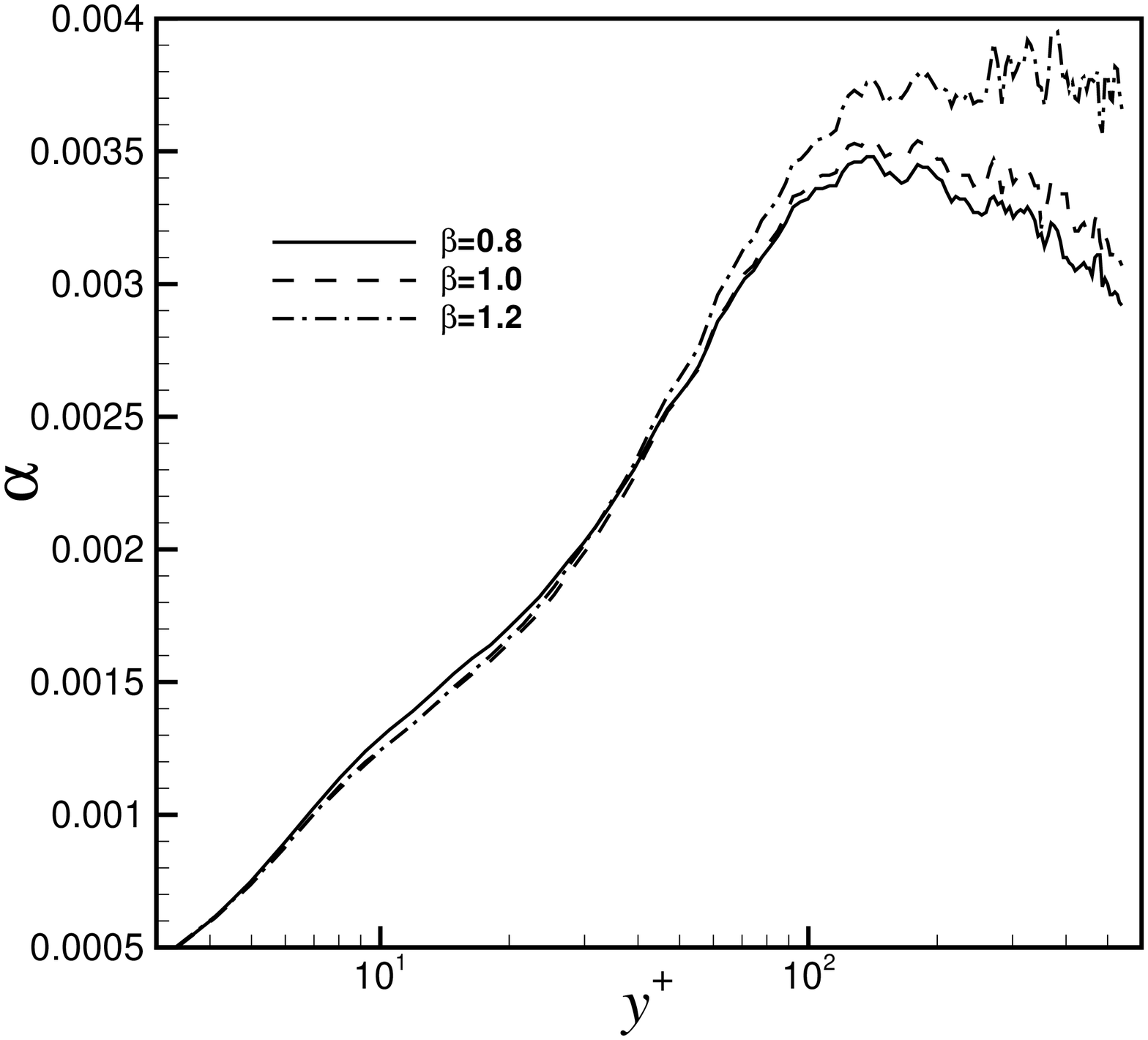},width=0.98\linewidth}
  \end{center}\end{minipage}\\
\begin{minipage}{0.48\linewidth} \begin{center} (a) \end{center} \end{minipage}
\begin{minipage}{0.48\linewidth} \begin{center} (b) \end{center} \end{minipage}
  \caption{Variation of $\alpha$ with distance from the wall in (a)
  global, and (b) wall units.}
\label{alpha_y_yplus}
\end{center}
\end{figure}

Figure \ref{alpha_y_yplus} shows the variation of $\alpha$ with the
distance from the wall in both global and wall coordinates for
$\beta=0.8,1.0,$ and $1.2$. As demonstrated in Figure
\ref{alpha_y_yplus}(b), $\alpha$ values experience a sharp increase in
the vicinity of the wall up to $y^+=100$. This region of sharp
increase in the value of $\alpha$ contains both the viscous sublayer
and the buffer layer. Diminishing values of $\alpha$ is observed as
one approaches the wall. This is consistent with theoretical
expectations that the NS equations ought to be recovered in the
laminar layer at the wall. The profile of $\alpha$ in the vicinity of
the wall shows minimal dependency on $\beta$. Away from the wall and
beyond $y^+=100$, $\alpha$ shows little variation across the
channel. One can argue that the dynamic LANS-$\alpha$ equations in
this case divides the flow into two distinct regions: a near wall
region that includes both the viscous sublayer and the buffer layer
where $\alpha$ is a function of the distance from the wall, and a
constant $\alpha$ region which includes the log layer and the outer
layer.  In the near wall region $\alpha$ keeps an almost log relation
with the distance from the wall in wall units. In summary, one can
argue that in wall bounded flows, the isotropic LANS-$\alpha$
calculations could be used with a constant $\alpha$ beyond $y^+=100$
and with a logarithmic relation in the near region. This projection
requires further investigation in LANS-$\alpha$ calculations.

\begin{figure}[hbt]
\begin{center}
\epsfig{file=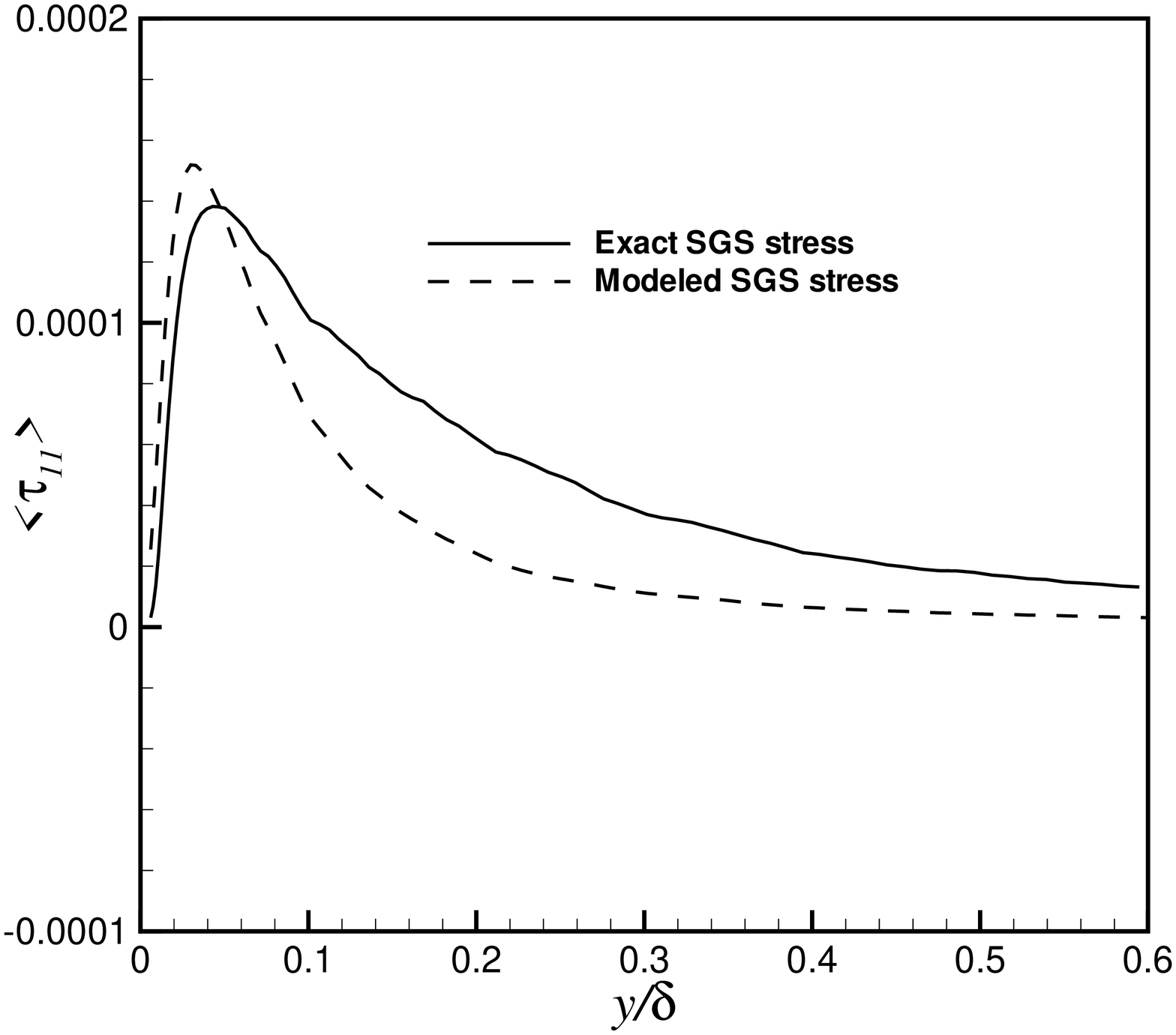}, width=0.6\linewidth}
\caption{The averaged subgrid scale normal stress $<\tau_{11}>$ in
  global units.} 
\label{sgs_t11}
\end{center}
\end{figure}

\begin{figure}[hbt]
\begin{center}
\epsfig{file=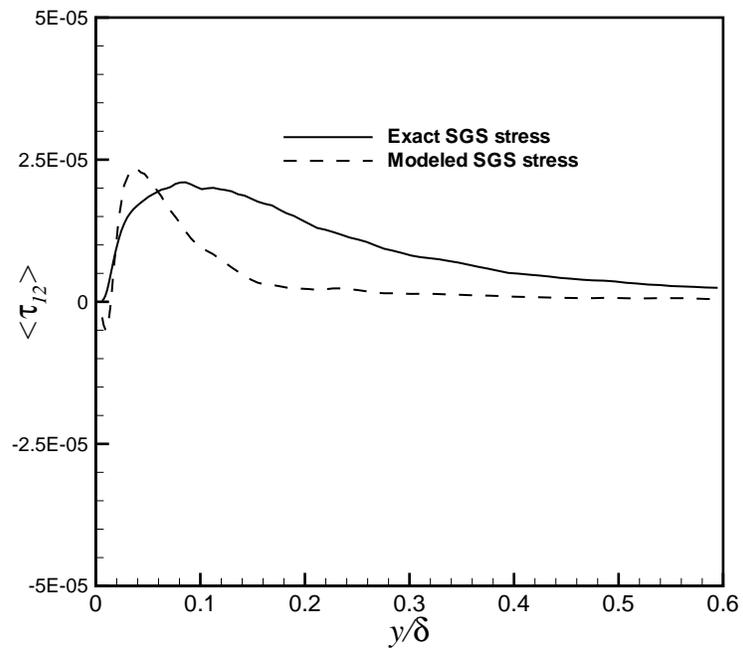}, width=0.6\linewidth}
\caption{The averaged subgrid scale shear stress $<\tau_{12}>$ in
  global units.} 
\label{sgs_t12}
\end{center}
\end{figure}

\begin{figure}[hbt]
\begin{center}
\epsfig{file=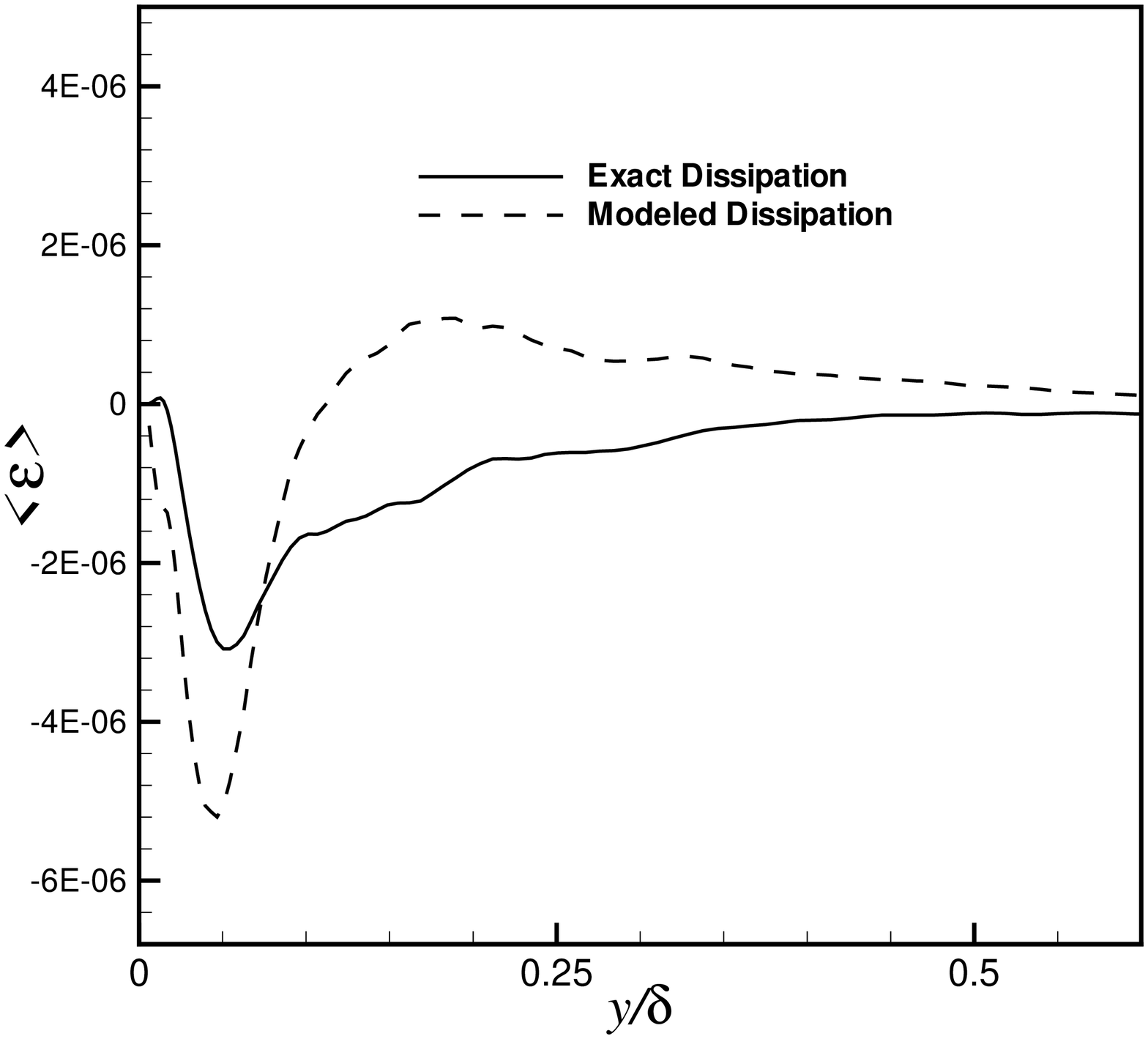}, width=0.6\linewidth}
\caption{The averaged dissipation in global units.}
\label{dissipation}
\end{center}
\end{figure}

Similar to the dynamic LES, one expects that the accuracy of the
dynamic LANS-$\alpha$ model to depend on its capability of accurately
modeling the subgrid scale stresses. The modeled and the exact SGS
stresses in this flow are shown in Figure \ref{sgs_t11} for the
isotropic component $<\tau_{11}>$ and Figure \ref{sgs_t12} for the
shear stress component $<\tau_{12}>$, where $<\cdot>$ stands for
averaging in streamwise and spanwise directions. The general
trend of the SGS stresses are captured in the dynamic LANS-$\alpha$
model without any ad hoc damping function. Good agreement
between the modeled and the exact SGS stresses in the near wall region
are observed. The SGS stresses vanish at the wall and in the middle of
the channel with a maximum value within the wall layer. 
The exact and modeled dissipation $<\varepsilon>$ are compared in
Figure \ref{dissipation}. Both the SGS stresses and the modeled
dissipations are effectively insensitive to variation in $\beta$. 


\section{Conclusions \label{sec:Conclusions}}
A dynamic LANS-$\alpha$ model is proposed where the variation in the
parameter $\alpha$ in the direction of anisotropy is determined in a
self-consistent way from data contained in the simulation itself. The
model results in a nonlinear equation for $\alpha$.  Numerical
experiments for decaying and forced homogenous isotropic turbulence
are performed using the dynamic LANS-$\alpha$ model. The simulation
results in both cases show an improvement over the LANS-$\alpha$
simulations with a fixed $\alpha$. 

{\it A priori} test of the dynamic LANS-$\alpha$ model in a channel
flow is carried out, where good agreement between the dynamic
LANS-$\alpha$ predictions and the DNS data is observed. The parameter
$\alpha$ is found to rapidly change in the wall normal direction in
the vicinity of the wall. Near the solid wall, the length scale
$\alpha$ shows a logarithmic dependence on the wall normal direction
in wall units. Away from the wall, and in the middle of the channel,
$\alpha$ approaches an essentially constant value. As a result, the
turbulent flow is divided into two regions: a constant $\alpha$ region
away from the wall and a near wall region. In the near wall region,
$\alpha$ keeps an almost logarithmic relation with the distance from
the wall. Consequently, one can argue that in wall bounded flows, the
isotropic LANS-$\alpha$ calculations could be used with a constant
$\alpha$ beyond $y^+=100$ and with a logarithmic relation in the near
region. These results indicates a promising application of the dynamic
LANS-$\alpha$ model in wall bounded turbulent flow simulations.

\section{Acknowledgement}
The research in this paper was partially supported by the AFOSR
contract F49620-02-1-0176.  The authors would like to thank B. Kosovic
for his initial help in the derivation of the dynamic model and
T. Lund for helpful discussions.  The DNS data of the channel flow was
generously provided by R. Moser and J. Jim\'enez.


\end{document}